# Compression of Femtosecond Pulses in a wide Wavelength Range Using a Large Mode Area Tapered Fiber

M. Rehan[1,*], G. Kumar[1], V. Rastogi[1], D. A. Korobko[2], and A. A. Sysolyatin[2, 3]

[1]Department of Physics, Indian Institute of Technology Roorkee, Roorkee-247667, U.K., India.
[2]Ulyanovsk State University, L. Tolstoy Str., 42, 432017 Ulyanovsk, Russia.
[3]General Physics Institute, Russian Academy of Sciences, 38 Vavilov Str., 119333, Moscow, Russia.
*rehan.dph2016@iitr.ac.in

## Abstract

We report design of a tapered fiber that can be used for compression of pulses at different central wavelengths. The proposed fiber is a 3-layer W-type large-mode-area fiber, which has been tapered to transform the mode area from 1700 μm$^2$ to 900 μm$^2$. We determine the exact length of the maximum pulse compression and numerically demonstrate the compression of 250 *fs*, 100 kW peak power input pulse to 15 *fs*, 850 kW at 1.55 μm wavelength and 250 *fs*, 120 kW peak power input pulses to 28 *fs*, 700 kW and to 46 *fs*, 500 kW at 1.8 μm and 2 μm wavelengths, respectively. Such a fiber can find wide ranging applications including in communication, spectroscopy and medicine.

## 1. Introduction

The generation and propagation of high peak power ultra-short pulses through a short distance of an optical fiber at different wavelengths have potential in a variety of high-power applications [1-4]. For example, high peak power ultra-short pulses, at 1.55 μm wavelength is used in optical communication system [5], pulses of 1.8 μm wavelength are useful in spectroscopy and laser ablation [6, 7], while pulses at 2 μm wavelength are used in medical treatments [7, 8]. Various investigations demonstrate that these high peak power ultra-short pulses can be generated using fiber lasers and amplifiers but fiber lasers suffer from complexity of the system and commercial feasibility [9]. In addition, nonlinear effects and thermal tolerance limit the output power of the fiber lasers. To overcome these difficulties, the optical pulse compression is another alternative [10-12]. In 1980, Mollenauer et *al.* were first who observed this phenomenon of pulse compression [13]. Later on, they have also reported the extreme picosecond pulse narrowing by using soliton effect in single-mode optical fibers [14]. In general, pulse compression is the consequence of the interplay between spectral broadening due to dispersion and self-focusing induced by nonlinear effects [15]. At high power, above 1.3 μm wavelength, in fused silica fiber, the inherent anomalous dispersion interacts with self-phase modulation and leads to the optical compression of the pulse [4]. The pulse compression using axially varying fibers is one of the trending approaches. Axially varying fibers like dispersion decreasing fibers, dispersion shifted fibers and tapered fibers are being used for the purpose of pulse propagation, supercontinuum generation and pulse compression [15-19]. A. Andrianov et *al.* reported the generation of widely tunable few-cycle 20-25 fs short duration pulses in the range of 1.6-2.1 μm by compression of femtosecond pulses using an all-fiber Er-doped laser source [20]. But, they have dealt with low peak power femtosecond pulse. In recent past, J. Yuan et *al.* proposed an inversely tapered silicon ridge waveguide to compress the picosecond pulses in mid infrared region [21]. They have realized exponential decreasing dispersion along the length of the fiber. The arrangement for the optical pulse compression is simple but necessitates adjusting many parameters such as length of the fiber, mode field diameter, input power, input pulse width and dispersion to achieve pulse compression. Recently, M. Gehbhardt et *al.* have reported generation of ultra-short pulses of 13 *fs* duration and 1.4 GW peak power around 2 µm wavelength from a thulium-doped fiber laser by using a gas filled anti-resonant hollow core fiber [22]. However, it involves handling of gas filled hollow core fiber. C. Gaida et *al.* reported self-compression of pulses in a solid-core fused silica fiber. They have attained pulses of 38 *fs* duration with average power of 24.6 W around 2 µm wavelength [23].

In this paper, we propose a large mode area tapered fiber for optical pulse compression in a wide wavelength range. Our recent work shows direction dependent propagation of femtosecond pulses in a tapered large-mode-area fiber [24]. We study the influence of the mode area variation with the length of the fiber and determine the precise length of the maximum pulse compression at different wavelengths. We numerically demonstrate the compression of femtosecond-Gaussian pulse of 250 *fs* duration with peak power of 100 kW at 1.55 µm wavelength through the 40 cm length of fiber and obtain the output pulse of 850 kW peak power with pulse duration of 15 *fs*. We show the compression of 250 *fs*, 120 kW peak power pulse by using 30 cm length of the same fiber at 1.8 µm wavelength to 700 kW, 28 *fs* duration pulses. In addition, we numerically demonstrate compression of 120 kW, 250 *fs* Gaussian pulse and achieve 500 kW, 46 *fs* duration output pulse at 2 µm wavelength. Compression of such high peak power *fs* pulses of different center wavelengths by using of the same fiber make the fiber design versatile and interesting for imaging and material processing applications [25, 26].

## 2. Fiber Design

To accommodate high peak power while maintaining good beam quality, we use a large-mode-area fiber based on higher order mode discrimination [27]. The nonlinear coefficient in a fiber is related to the mode effective area by the following relation

$$\gamma = \frac{\omega_o n_2}{c A_{eff}} \tag{1}$$

where, $\gamma$, $\omega_o$, $n_2$, $c$ and $A_{eff}$ are represent the nonlinear coefficient, central frequency, nonlinear refractive index, speed of light, and mode effective area of the fiber, respectively. To satisfy above mentioned requirements, we have taken the large mode area three layered structure fiber proposed by Babita et *al.* [28] and down tapered it such that the width ratio of all the three layers remains unchanged. Here, the mode area varies from 1700 µm² to 900 µm² at 1.55 µm wavelength. The normalized intensity and refractive index profiles with radial position and schematic of down tapered fiber are shown in figure 1(a) and 1(b), respectively. Here, the core of the fiber is made of Ge-doped silica ($n_c$) and depressed and outer claddings are that of F-doped silica ($n_d$) and pure silica ($n_s$), respectively. The values of various parameters are chosen as core radius ($a$) = 30 µm, width of the depressed cladding ($b$) = 15 µm, width of outermost layer ($c$) = 17.5 µm, relative index difference of core with respect to outer most silica layer is $\Delta_1$ = 0.03%, and that of depressed cladding is $\Delta_2$ = 0.08%. $\Delta_1$ and $\Delta_2$ are taken such that the fiber supports $LP_{01}$ and $LP_{11}$ modes only and all the other higher order modes are stripped off. We have used transfer matrix method (TMM) to analyze the modal characteristics of the fiber [29].

The pulse compression without pulse breaking through such a fiber is obtained by appropriate tapering of the fiber. Tapering of the fiber leads to variations in effective mode area ($A_{eff}$) and effective index ($n_{eff}$) of the fiber. The changes in $n_{eff}$ leads to the change in the dispersion ($D$). The mathematical forms of $A_{eff}$ is given as

$$A_{eff} = \frac{2\pi[\int_0^\infty |\phi_r|^2 rdr]^2}{\int_0^\infty |\phi_r|^4 rdr} \qquad (2)$$

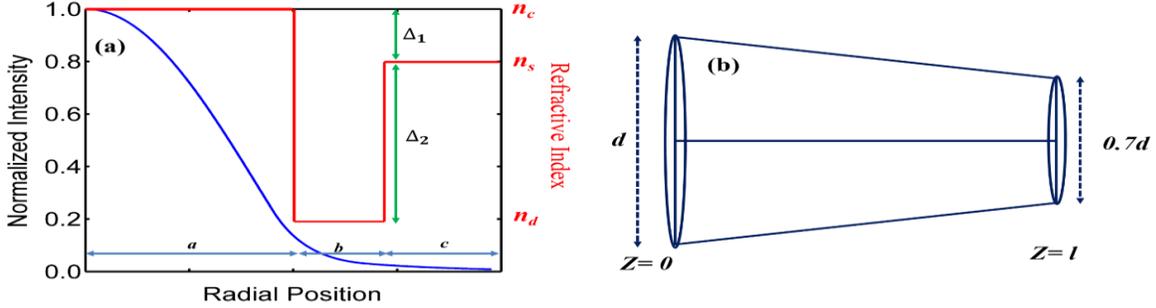

Figure 1: (a) Refractive index profiles of three-layered structure fiber along with the normalized intensity profile of the LP$_{01}$ mode and (b) Schematic of down tapered fiber. Here, $d$ and $l$ represent the diameter and length of the fiber, respectively.

And, dispersion coefficient is given by

$$D = -\frac{\lambda}{c}\frac{\partial^2 n_{eff}}{\partial \lambda^2} \qquad (3)$$

where, $\phi_r(r)$, $c$, and $\lambda$ represent the radial part of the field, speed of light, and central wavelength, respectively. Hence, dispersion and nonlinearity vary along the length of the fiber with this axial variation i.e. tapering of the fiber. This tapering reduces the mode area over the length of the fiber and increases the nonlinearity without significantly affecting the dispersion. This leads to compression of the pulses.

## 3. Theory

In order to realize the propagation of ultra-short pulses of high peak power through the optical fiber, it is important to include dispersion, nonlinear effects, and losses in the fiber. The equation which governs the aforementioned propagation of pulses through optical fiber is the nonlinear Schrodinger equation (NLSE) [4, 30].

$$\frac{\partial A(t,z)}{\partial z} = \left(-i\frac{\beta_2}{2}\frac{\partial^2}{\partial T^2} + \frac{\beta_3}{6}\frac{\partial^2}{\partial T^2} - \frac{\alpha}{2} + i\gamma\left(|A|^2 + \frac{i}{\omega_0}\frac{1}{A}\frac{\partial}{\partial T}(|A|^2 A) - T_R\frac{\partial |A|^2}{\partial T}\right)\right)A \qquad (4)$$

where, $A(t, z)$ represents the slowly varying envelope of pulse amplitude, $\alpha$ is the loss coefficient of the fiber. The parameters $\omega_o$ and $T_R$ are the central frequency and Raman time constant. Here, $\beta_2$ and $\beta_3$ are the dispersions of second and third order, respectively. $T$ represents scaled time, which is given by

$$T = t - \frac{z}{v_g} \qquad (5)$$

And, the second order dispersion is given by

$$\beta_2 = -D\frac{\lambda^2}{2\pi c} \qquad (6)$$

where, $v_g$ and $D$ represent the group velocity of pulse envelope and dispersion coefficient, respectively. On tapering the fiber, $D$ varies along the length of the fiber. Hence, the second order dispersion coefficient $\beta_2$ also changes, accordingly. The variation of third order dispersion coefficient $\beta_3$ is very small so that we have chosen the constant values of 0.104 ps³/km, 0.157 ps³/km, and 0.149 ps³/km at wavelengths 1.55 μm, 1.8 μm, and 2 μm, respectively. We have used split-step Fourier method (SSFM) to solve the above NLSE [4]. Here, we have tapered the fiber to optimize nonlinearity and dispersion in order to achieve pulse compression.

## 4. Results and Discussion

We have studied the propagation of a femtosecond Gaussian pulse through proposed LMA tapered fiber. The mathematical form of input pulse is given by

$$A(t, z = 0) = A_o \exp(-t^2/2t_0^2) \tag{7}$$

Where, $A_o(= \sqrt{P_o})$ represents the pulse amplitude, $P_o$ stands for the input power, and $t_o$ is the input pulse width. We have down-tapered the fiber and optimized the input parameters in such a way that nonlinear effects, mainly self-phase modulation (SPM) dominates over dispersion and pulse compression is achieved. In our study the effect of stimulated Raman scattering (SRS) is negligible, since the femtosecond pulses and the generally Raman vibrations are also in the order of hundreds of femtoseconds [31, 32]. Other than this, the effective mode area of our fiber is in the order of $10^3$ μm² which increase the Raman threshold [4].

### 4.1. (a) Pulse Compression at 1.55 μm wavelength

To demonstrate the compression of the Gaussian pulse at 1.55 μm, we have solved NLSE for propagation of an input pulse of 250 *fs* duration with peak power of 100 kW through 40 cm length of the proposed fiber. The temporal shapes of the input and output pulses are shown in figure 2(a) by blue and red colors, respectively which clearly shows the compression of the pulse. The peak power of output pulse is 850 kW and pulse duration is 15 *fs*. The compression factor of approximately 17 has been obtained. Corresponding contour plot and the pulse evolution along the length of the fiber are shown in figure 2(b) and 2(c), respectively.

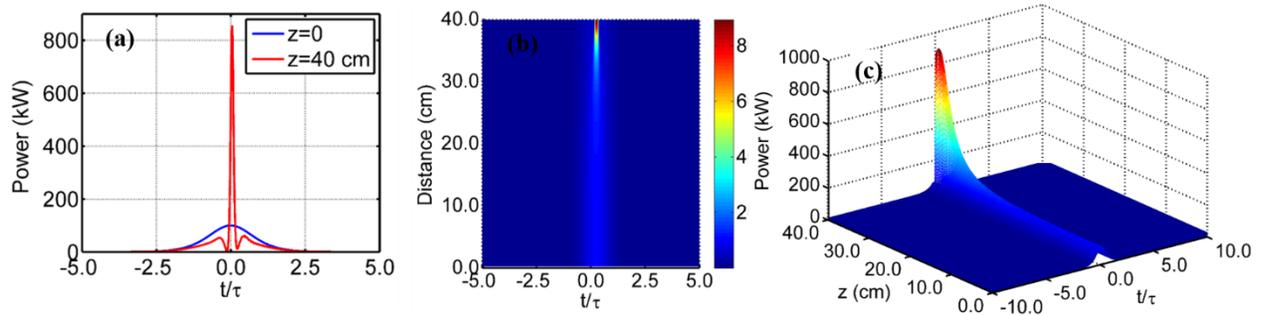

Figure 2. (a) The temporal profiles of input/output pulses, (b) the corresponding contour plot, and (c) the pulse evolution along the 40 cm length of LMA tapered fiber at 1.55 μm wavelength.

To understand the underlying physics, the dispersion ($D$) and nonlinear coefficients ($\gamma$) have been plotted with the length of the fiber in figure 3(a) and figure 3(b), respectively. Figure 3(a) and 3(b) show that the dispersion and nonlinear coefficient are increasing along the length of the fiber but the increase in nonlinear coefficient is larger than that in dispersion. Consequently, SPM dominates over GVD along the length of the fiber and compression of the pulse observed. There are various parameters such as input pulse width, input peak power, fiber length and dispersion that need to be optimized to achieve the optical pulse compression. For the chosen fiber parameters, the pulse width has been varied for a constant input peak power to obtain the maximum compression of the optical pulse. Similarly, for a constant input pulse width the input peak power has been varied to optimize the peak power of input pulse.

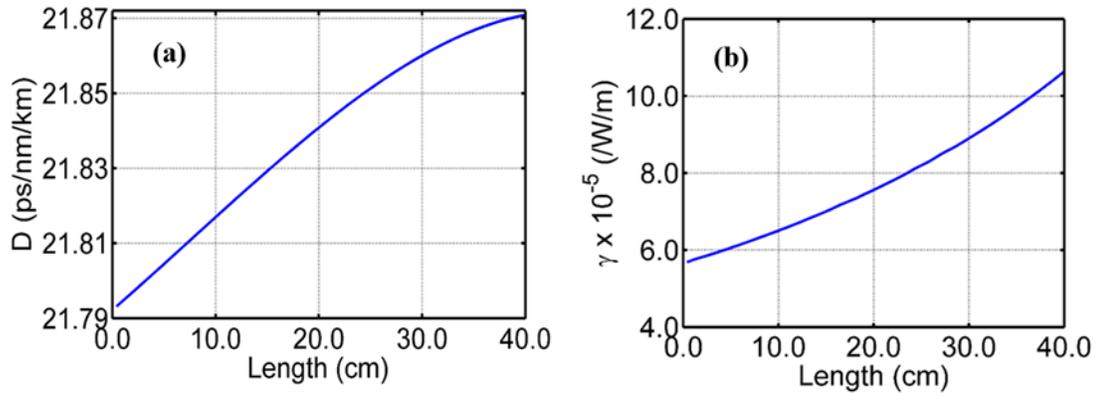

Figure 3: Variation of (a) dispersion coefficient (*D*), and (b) nonlinear coefficient ($\gamma$) with the length of the fiber at 1.55 µm wavelength.

### (b) Effect of input pulse width and peak power

The pulse width of input pulse has been varied from 100 *fs* to 300 *fs* in steps of 5 *fs* and the pulse propagation over 40 cm length of the LMA tapered fiber has been studied. The input peak power has been chosen as 100 kW. The ratio of output to input pulse widths has been plotted with respect to the length of the fiber as shown in figure 4 (a).

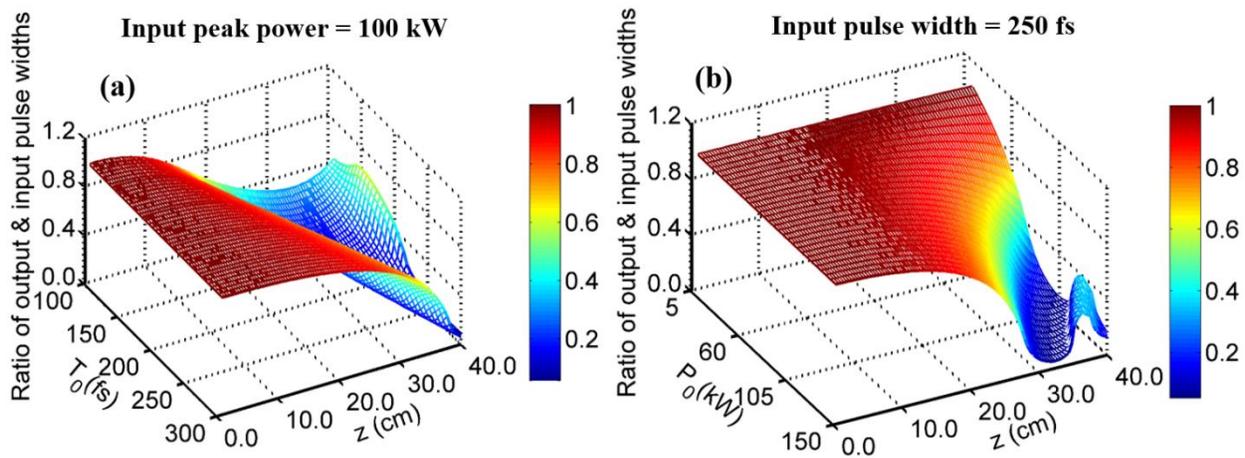

Figure 4. Raito of output and input pulse widths with 40 cm length of the fiber, for (a) pulse widths varying from 100 *fs* to 300 *fs* with peak power of 100 kW, (b) peak power is varied from 5 kW to 150 kW while keeping the pulse width 250 *fs*. Here, $T_0$, $P_0$, and z represent the input pulse width, input peak power and length of the fiber.

The plot shows the pulse compression for the chosen range of pulse width. The compression of the pulse increases with the increase in the input pulse width and the maximum compression of the pulse has been observed around 250 *fs* pulse width. Hence, we have chosen value of pulse width as 250 *fs* in our further calculations.

In a similar way, we have fixed the value of input pulse width as 250 fs and varied the peak power of input pulse from 5 kW to 150 kW for 40 cm length of the proposed fiber. We have plotted the ratio of output to input FWHM with the length of the fiber for different values of input peak powers as shown in figure 4(b). It has been observed that with the increase in the peak power, the pulse compression increases and the maximum compression is achieved at 100 kW. On further increase in peak power, the pulse compresses till 30 cm of length and beyond that it starts dispersing periodically.

### *4. 2. (a) Pulse Compression at 1.8 µm wavelength*

Ultra-short pulses with high peak power around 1.8 µm wavelength can be used in spectroscopy [2, 4]. When an ultra-short Gaussian pulse of 250 fs pulse width with 120 kW peak power propagates through the 30 cm length of proposed fiber, the pulse has been compressed to 28 fs, 700 kW. The corresponding input-output (blue-red) pulses are shown in figure 5(a).

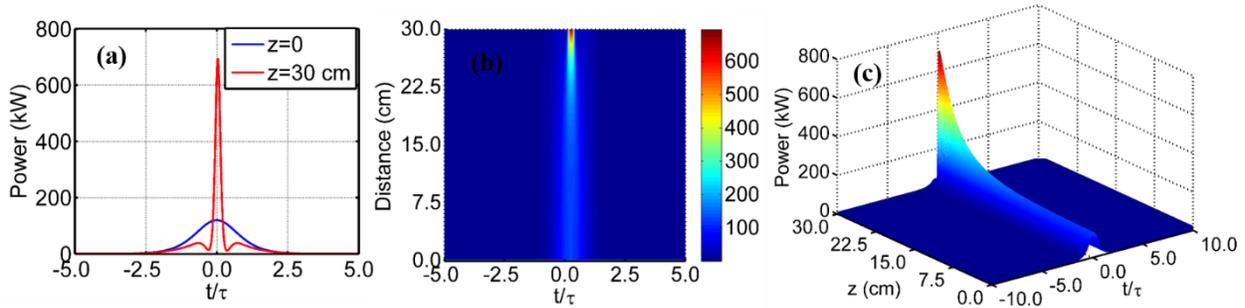

Figure 5. (a) The temporal profiles of input/output pulses, (b) the corresponding contour plot, and (c) the pulse evolution along the 30 cm length of LMA tapered fiber at 1.8 µm wavelength.

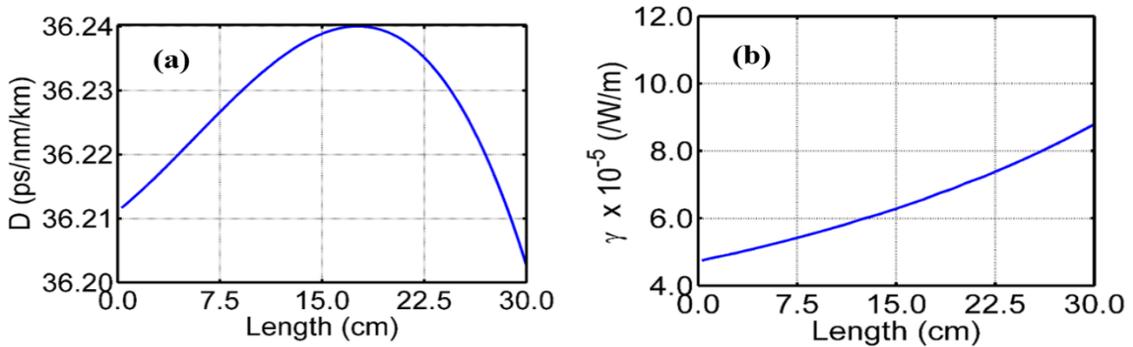

Figure 6: Variation of (a) dispersion coefficient ($D$), and (b) nonlinear coefficient ($\gamma$) with the length of the fiber at 1.8 µm wavelength.

Here, the compression of the pulse can be seen clearly. The compression factor of 9 has been obtained. The contour plot and the pulse evolution through the length of the fiber are shown in figs. 5(b) and 5(c), respectively. The results show that the SPM dominating GVD along the length of the fiber. It can be understand by looking the variation in dispersion and nonlinear coefficient with the length of the fiber as plotted in figs. 6(a)-(b). The plot of dispersion shows that the dispersion coefficient increases till 17 cm length and then decreases drastically in the remaining length of the fiber. On the other hand, plot of nonlinear coefficient increasing throughout the length of the fiber. Hence, SPM dominates over GVD and the compression of the pulse is observed along the length of the fiber.

On comparing figure 3 and figure 6, it can be clearly seen that the value of dispersion coefficient is higher in case of 1.8 µm wavelength while the nonlinear coefficient is lower in case of 1.8 µm wavelength. Hence, the strength of dominance of SPM over GVD is lower in case of 1.8 µm. Therefore, the pulse compression factor is lesser but the achieved compression of the pulses is reasonable to use in high power laser applications. On keeping the peak power constant, we have varied width of input pulse to find its value for maximum compression. Similarly, we have fixed the input pulse width and varied input peak power.

*(b) Effect of input pulse width and peak power*

For 120 kW chosen value of the input peak power, we have varied the input pulse width from 100 *fs* to 300 *fs* and the pulse propagation over 30 cm length of the proposed LMA tapered fiber has been studied. For different values of pulse width, the ratio of output and input pulse widths with the length of the fiber is shown in figure 7 (a).

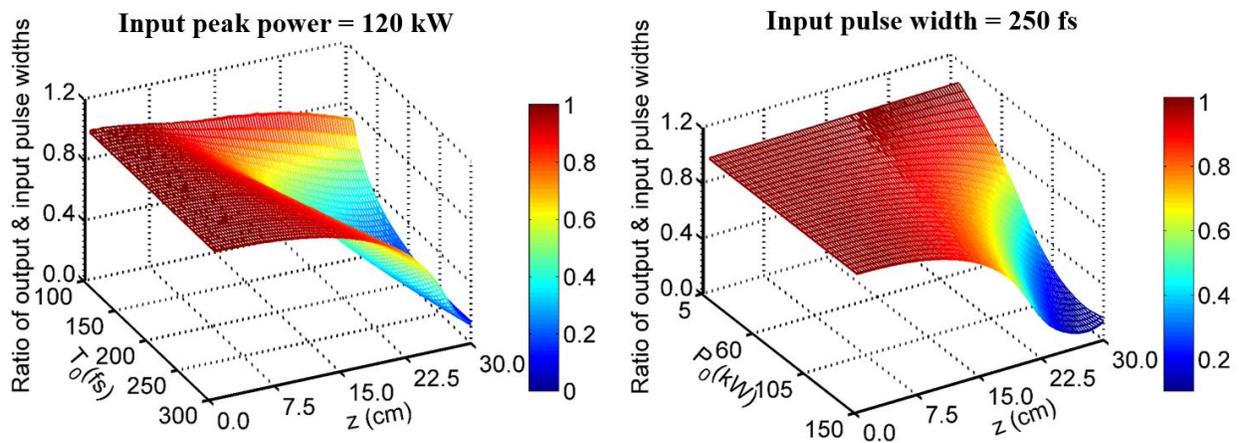

Figure 7. Raito of output to input FWHM with 30 cm length of the fiber, for (a) pulse width 100 *fs* to 300 *fs* with constant 120 kW peak power, (b) peak power is varied from 5 kW to 150 kW with 250 *fs* fixed pulse duration. Here, $T_0$, $P_0$, and z represent the input pulse width, input peak power and length of the fiber.

This shows, on increasing the input pulse width, compression of the pulse increases and the maximum compression of the pulse has been observed for 250 *fs* input pulse duration. For 250 *fs* pulse duration, we have varied the input peak power from 5 kW to 150 kW for the same length of fiber and the ratio of output and input pulse widths with respect to the length of the fiber is shown in fig 7(b). This can be seen that for input peak power around 5 kW, the output pulse is similar to that of input while on further increasing the input peak power pulse compression is observed. The maximum pulse compression has been achieved around 120 kW input peak power due to dominance of SPM over GVD.

*4. 3. (a) Pulse Compression at 2 µm wavelength*

Ultra-short pulses of high peak power at 2 µm wavelength are useful in high power medical applications [3]. We have studied the propagation of an input Gaussian pulse of 250 *fs* duration with peak power of 120 kW through 30 cm length of the proposed LMA tapered fiber. The pulse of 500 kW peak power and 46 *fs* duration is obtained. The pulse width of output pulse reduced approximately 6 times that of the input pulse. The temporal shapes of the input and output pulses along with contour plot, and the pulse evolution over the length of the fiber have shown in figs. 8(a)-(c).

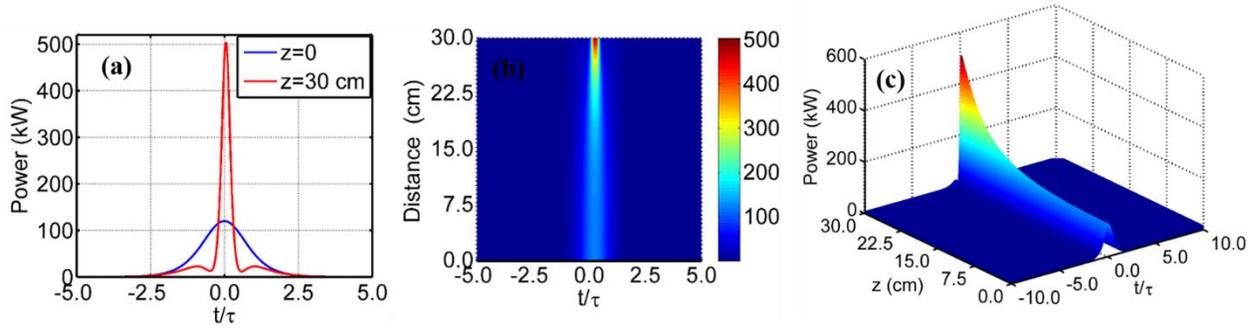

Figure 8. (a) The temporal profiles of input/output pulses, (b) the corresponding contour plot, and (c) the pulse evolution along the 30 cm length of LMA tapered fiber at 2 µm wavelength.

The three layered structure LMA fiber has been down tapered so that the dispersion and nonlinear coefficients are varying along the length of the fiber as shown in figure 9(a)-(b). The plots show that the dispersion coefficient is almost constant till 15 cm length and then starts decreasing along the length of the fiber and nonlinearity increasing along the length of the fiber. Hence, pulse compression is observed due to the interplay between SPM and GVD.

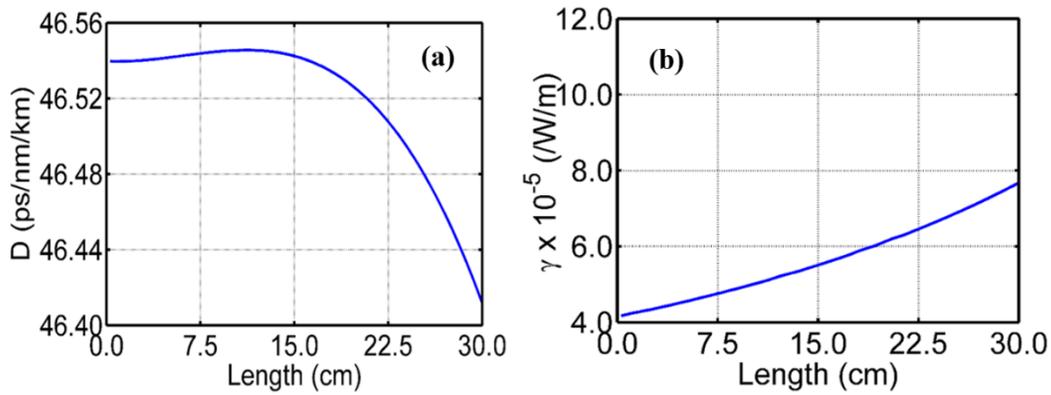

Figure 9: Variation of (a) dispersion, and (b) nonlinear coefficients with the length of the fiber at 2 µm wavelength.

From figure 3, figure 6, and figure 9, this can be clearly observed that value of dispersion coefficient is higher at higher wavelength vice versa and case is reversed for nonlinear coefficient. The strength of the nonlinearity is greater in case of 1.55 µm wavelength. Also, the loss coefficient ($\alpha$) is 0.2 dB/km at 1.55 µm wavelength is smaller than that of 1.3 dB/km and 2 dB/km at 1.8 µm and 2 µm wavelengths, respectively, for fused silica glass fiber [33]. Therefore, the compression factor is maximum at 1.55 µm then for 1.8 µm and after that for 2 µm wavelength. The input parameters like input pulse width and input peak power need to be optimized to obtain the significant optical pulse compression.

*(b) Effect of input pulse width and peak power*

The input pulse duration has been varied from 100 *fs* to 300 *fs* and the pulse propagation over 30 cm length of the proposed LMA tapered fiber has been studied for a chosen value of 120 kW input peak power. For the selected range of input pulse width, the ratio of output and input pulse widths has been plotted with respect to the length of the fiber, which is shown in the figure 10 (a). We have observed dispersion of the pulse for input pulse width around 100 *fs* while on increasing the input pulse width pulse similar to the input is realized. On further increasing the input pulse width compression of the pulse is attained. We have observed maximum pulse compression around 250 *fs* input pulse width. Similarly, we have varied input peak power from 5 kW to 150 kW for 250 *fs* input pulse width in the 30 cm length of the fiber and plotted ratio of output and input pulse widths with the length of the fiber for all the values of input peak power as shown in figure 10(b).

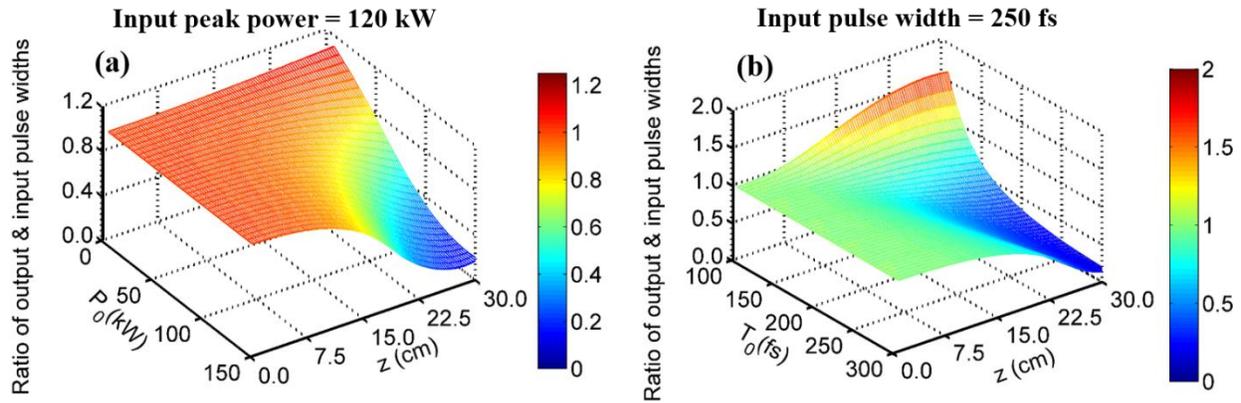

Figure 10. Raito of output to input FWHM with the 30 cm length of the fiber, for (a) pulse width 100 *fs* to 300 *fs* with constant 120 kW peak power, (b) peak power is varied from 5 kW to 150 kW with 250 *fs* fixed pulse duration. Here, $T_0$, $P_0$, and z represent the input pulse width, input peak power and length of the fiber.

This is observed that the ratio of output and input pulse widths is lower around 150 kW and lowest at 120 kW input peak power due to dominance SPM over GVD. Hence, the chosen value of input peak power for best compression is 120 kW in our further calculations.

## 5. Conclusion

We have proposed a new design of a three-layer W-type structure large mode area down tapered fiber and numerically demonstrated compression of 250 *fs* duration pulse at different wavelengths by means of the same fiber. The exact length of the maximum pulse compression is determined and the influence of the mode area variation with the length of the fiber has been studied. For 1.55 μm wavelength, a 250 *fs*, 100 kW pulse compressed to 15 *fs*, 850 kW while propagated over 40 cm length of the proposed fiber. These ultra-short pulses can be utilized in ultra-high data rate optical communication. For 1.8 μm wavelength, a 250 *fs*, 120 kW pulse turn out to be 28 *fs*, 700 kW while transmitted from 30 cm length of the proposed tapered fiber. For 2 μm wavelength, a pulse of 250 *fs*, 120 kW is become 46 *fs*, 500 kW while delivered through 30 cm length of the tapered fiber. These pulses in the eye safe region can be used in spectroscopy, remote sensing and medical treatments. Hence, our fiber design can be a good candidate to produce high peak power ultra-short pulses in a wide wavelength range for a number of high power applications.

*Acknowledgment:* This work has been partially supported through Department of Science and Technology New Delhi and Russian Science Foundation (DST-RSF) by Indo-Russian project on "Research and development of new optical fibers for applications in modern laser systems."